Tuning friction via electro-convoluted lipid-membrane boundary layers

Di Jin* and Jacob Klein

Dept. of Molecular Chemistry and Materials Science, Weizmann Institute of Science, Rehovot, Israel

**Abstract**

The recent discovery that sliding friction between surfaces bearing lipid bilayers is massively and reversibly modulated by a transverse electric field was attributed to electroporation and interbilayer bridging. Here we calculate the friction that arises from both the electroporation, due to closer contact between the lipid headgroups, and the bridging which topologically-forces the slip-plane to pass through the dissipative acyl tail-tail interface. Our results account quantitatively for the electro-modulated friction, indicating how such potentially-ubiquitous effects can be tuned.

1   **Introduction**

Lubrication is a vital process in physiological environments such as at the cartilage surface of a joint, where the low friction has been attributed to phospholipids, mainly phosphatidylcholines (PCs), forming boundary layers at the cartilage surface. Such lipid bilayers reduce frictional dissipation via the hydration lubrication mechanism: under compression, the PC lipid membranes remain intact and form bilayer stacks separated by hydration repulsion, which in the hydration lubrication mechanism enables easy slip at the highly-hydrated interface between opposing bilayers [1-12]. We recently discovered that hydration lubrication by such PC-lipid membranes can be massively modulated by electric fields [13]. In these experiments, lipid bilayers or liposomes were adsorbed to the molecularly-smooth mica and gold surfaces under normal stresses of O(10) atm. At a shearing velocity around 1 μm/s, it was discovered that a ca. 0.1V/nm transmembrane electric field massively increases the friction coefficient μ (= [force to slide]/load), from ca. 0.0005 to 0.1, an unprecedently drastic variation achieved by any electro-tuning mechanism (Table 1). This behavior is fully reversible, as the outstanding interbilayer lubricity is fully restored when the electric field is turned off. The current study aims to investigate the molecular origin of this friction switching behavior using molecular dynamics simulations.

MD simulations have long been applied to study friction of lipid bilayers though not in the presence of electric fields. Most of these studies focus on interleaflet friction [14-18]. Boţan et al. (2015) studied interbilayer friction with gel phase and liquid crystalline phase lipids, showing a strong correlation with the hydration level [19]. Their membranes were unsupported, making comparisons to experiments less ideal. Our current POPC (1-palmitoyl-2-oleoyl-sn-glycero-3-phosphocholine) lipid system is designed to closely mimic the SFB experimental setup, including mica and gold confining phases, with hydration levels determined from experiments. Initial configurations, based on our previous simulations [20], are either confined, unperturbed bilayers (no electric field) or the structurally-convoluted electroporated stacks equilibrated under an electric field. Based on our previous analysis, the structural convolution triggered by an electric field similar to the magnitude applied in the SFB experiment is found to be hydration-dependent: while at both the well-established minimum hydration level $n_w \approx 12$ water/lipid and the elevated hydration level of $n_w \approx 20$ water/lipid attributed to the water defects in the supported lipid bilayers, the well-known electroporation behavior is observed, in the latter case, a lipid bridge also forms, inter-connecting the two membranes that are otherwise separated by hydration repulsion. Upon removal of the electric field, the pore and bridge structures fully disappear, consistent with the reversibility of the friction changes in the experiment when toggling the electric field. Here, we will employ the three structures resulted from our former equilibrium analysis to elucidate the friction mechanisms correlated with the complex structural dynamics triggered by the electric field [21]. We shall see that the structural/topological convolution is indeed the origin of reduced lubrication under the electric field, with quantitative agreement between the shear stress revealed by the simulations and that measured in the experiments.

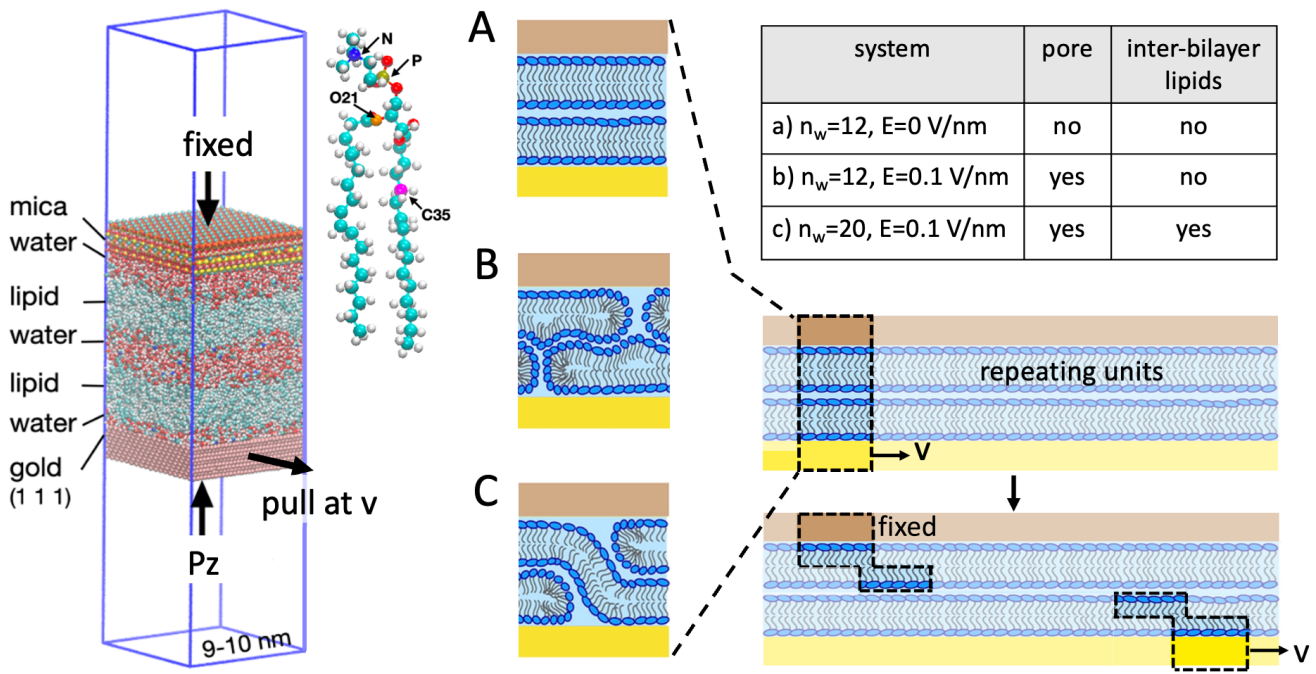

| system | pore | inter-bilayer lipids |
|---|---|---|
| a) $n_w$=12, E=0 V/nm | no | no |
| b) $n_w$=12, E=0.1 V/nm | yes | no |
| c) $n_w$=20, E=0.1 V/nm | yes | yes |

Figure 1. Left: Simulation setup of the confined double bilayer system. Mica: K (cyan), Si (yellow), Al (pink), O (red). Water: O (red), H (white). Lipid: C (cyan), H (white). Gold: pink. Right: Schematics of the three systems equilibrated under different hydration levels and electric fields. Due to periodic boundary conditions, each simulated system represents a repeating unit of the infinite planar membranes in confinement. The simulation is performed by pulling the gold slab at a designated velocity with harmonic potential while the mica surface is fixed in position. A constant force is applied to the gold slab, resulting in a pressure of $p_z$ = 10 atm. Structure b and c subject to an electric field of 0.1 V/nm imposed by the charge imbalance method (see methods).

## 2     Results and discussion

Figure 1 depicts the three confined bilayer structures that underwent shearing tests. The systems comprise two POPC bilayers that are hydrated and sandwiched between a mica slab and a gold slab (2 nm in thickness) and an area of 80-100 nm². The position of the mica surface is fixed, while the gold surface is pulled in the x direction using a harmonic potential, enabling the evaluation of the shear stress (see methods). The system is periodic in the x-y direction, thereby taking on the characteristics of an infinite planar structure. Consequently, the sliding of each monolayer block in our finite-sized systems closely approximates the sliding of the monolayers of a macroscopic double-bilayer structure in experiments. Structure A in Figure 1 represents the double bilayers compressed by the solid phases to the minimum hydration of $n_w$=12 in the absence of electric field, corresponding to the amount of hydration shell water molecules tenaciously attached to the zwitterionic lipid headgroups even under pressure [22-26]. Structure B represents system A porated under an electric field of ca. 0.1 V/nm. Structure C shows supported lipid membranes at elevated hydration levels due to the presence of water defects [27], as explained in our previous work, represented by a $n_w$=20 system, which leads to an additional feature of bridge formation under the same electric field. As we shall see later, shearing structure C leads to complex dynamics due to the presence of the bridge. The $n_w$=20, $E$=0 system (with similar unperturbed bilayers to structure A) is also included as a control for the analysis.

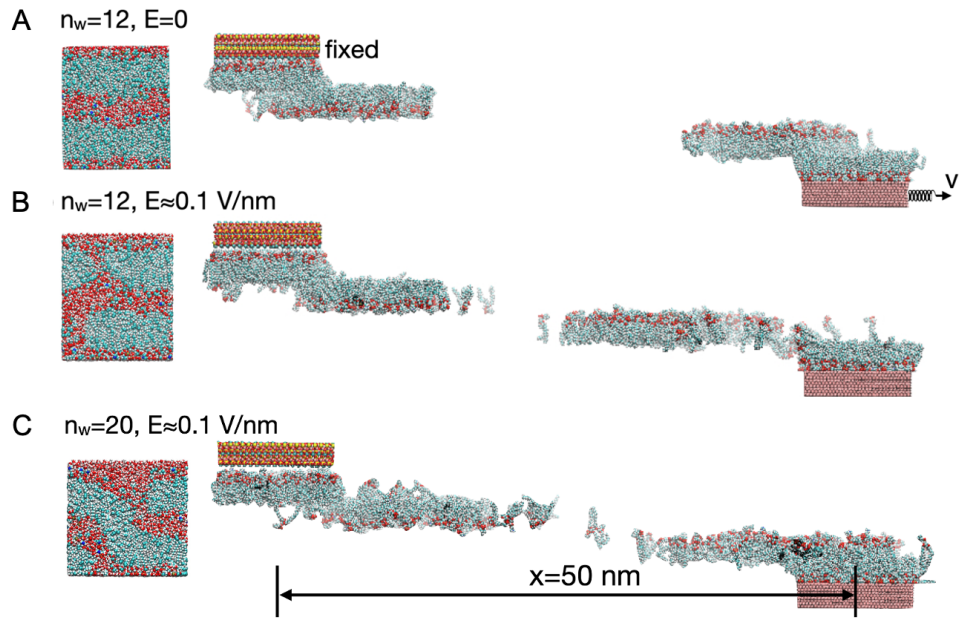

Figure 2. Simulated shearing of lipid bilayers at $v = 0.1$ m/s with the three systems A-C described in Figure 1. Periodic boundary condition is corrected to reveal the trajectories of atoms in the planar direction. A) Sliding mainly occurred at the midplane with some interleaflet sliding due to high shear stress resulted from the high velocities. B) Comparing to a, enhanced deformation is found in the midplane monolayers arising from higher friction force. C) Significant deformation at the midplane is observed under the presence of a lipid bridge and electropores.

**Enhanced intermembrane friction by electric field.**

Due to limited computational time, our simulation is limited to velocities of O(m/s), significantly higher than the SFB experimental velocity of O(μm/s). Nonetheless, these simulations shed strong light on the shearing dynamics and friction mechanism at experimental velocities, demonstrated by our quantitative analysis later below. Figure 2 shows the displacement of lipid molecules in different monolayers when sheared at the lowest velocity tested, 0.1 m/s, providing visual cues for comparison of lubrication properties among the structures. Structure A represents the membrane stack in absence of any electric field-induced structural convolution and experienced the least degree of membrane deformation. As described by the hydration lubrication scheme, the slip-plane during relative motion between bilayers is at the relatively fluid water phase at the inter-bilayer midplane, as that is the path of least frictional dissipation. Some intra-bilayer dissipation, i.e. at the tail-tail interface, also arises at the high sliding velocities simulated, as indicated in Figure 2 A (see theoretical analysis later below). In the case of the electroporated structure B even without any hydrophobic slip between the two bilayers, the lubrication is surprisingly poor as indicated by the significant deformation of the midplane leaflets. Structure C is both porated and with lipid bridging between the bilayers, and the deformation of the midplane monolayers is visibly apparent comparing to A. Although the significantly higher hydration level of structure C is expected to lead to improved lubrication based on the basic principles of hydration lubrication, the benefits are outweighed by modulation due to the structural variation.

The variation with time of shear stress at varying velocities for each of the three systems is shown in Figure 3. Steady-state shear stress monotonically increases with velocity in all structures. Figure 4 a shows the relation between measured steady-state shear stress as a function of sliding velocity. For all three systems, shear stress is approximately linearly dependent on $v$. The shear stress due to direct lipid-lipid interaction at the midplane $\sigma_{lipid}(v)$ is extracted (see later analysis), and also manifests a linear dependency (Figure 4 c). In addition, for both hydration levels and at a given velocity, the shear stress significantly increases in the presence of the electric field, and this trend seems to preserve at lower velocities, indicated by the extrapolation of $\sigma(v)$ in Figure 4 b. For both hydration levels, at $v\sim O(\mu m/s)$, the experiment relevant velocities, the shear stress of the $E$=0.1 V/nm systems is ca. $10^5$ Pa and that of the $E$=0 systems approaches zero within measurement error ($10^4$ Pa). Given the systems are under 10 atm of compression, this converts to friction coefficient which increases from an undetectable value to O(0.1) with the electric field switched on, consistent with the experimentally measured 0.0005 to 0.1 change in friction coefficient. Next, we exploit the advantage of the molecular details provided by the simulations and examine the dynamics of the electric-field-induced inter-membrane interactions to understand the drastically increased friction coefficient.

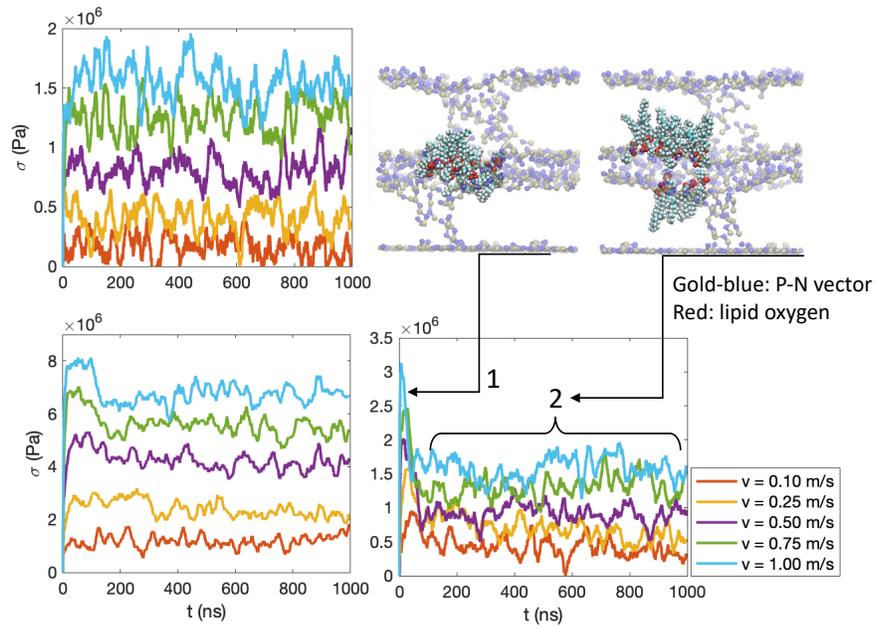

Figure 3. Time series of shear stress σ(*t*) measured from simulations of the lipid membrane systems with two hydration levels and two electric field conditions. With the structure $n_w$=20, $E$=0.1 V/nm, the shear stress undergoes two distinguishing mechanisms as labeled in the plot, respectively corresponding to before and after the dissociation of the bridge (For color coding of lipid atoms in snapshot see Figure 1).

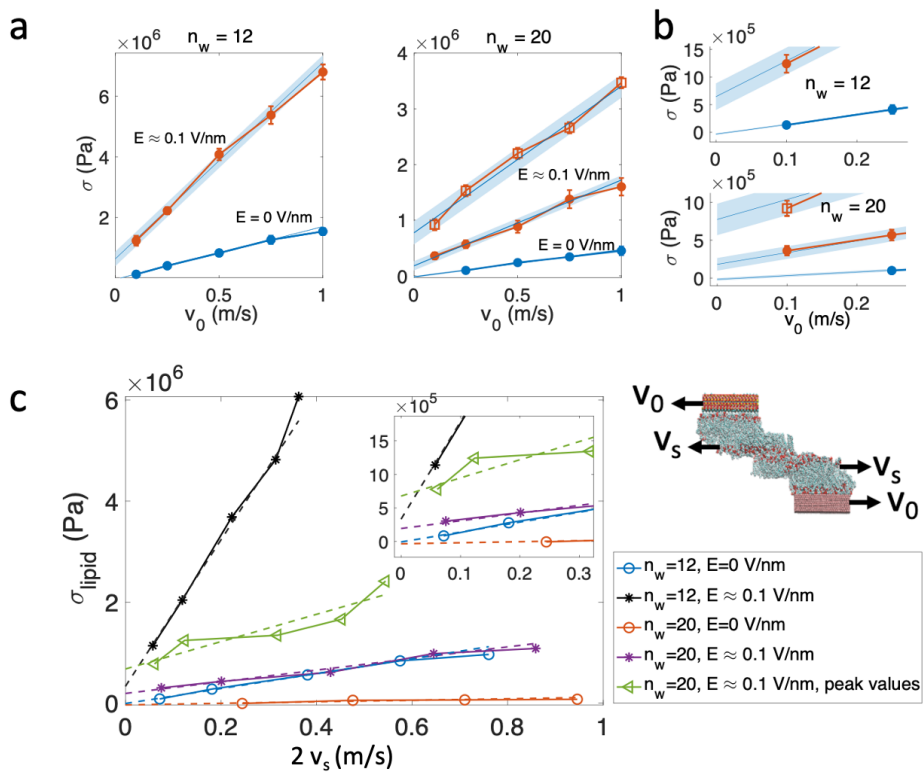

Figure 4. a) Steady-state shear stress $\sigma$ evaluated from the final 300 ns of the traces in Figure 3 as a function of pulling velocity $v_0$. For $n_w$=20, $E$=0.1 V/nm, the open symbols represent the peak $\sigma$ measured and the filled symbols represents the steady-state $\sigma$. b) Zoomed plot of Figure a to show extrapolation to the experimentally relevant velocities of O(μm/s). c) Interbilayer shear stress due to direct lipid-lipid interaction $\sigma_{lipid}$ estimated as a function of relative velocities between the two monolayers at the mid-plane. Dashed lines are linear fit for the extrapolation to the experimentally relevant velocities of O(μm/s).

**Enhanced structural linkage by lipid bridge.**

In Figure 3, a striking feature is found with structure C ($n_w$=20, $E$=0.1 V/nm), where all shear stress time series take an initial surge in magnitude and quickly reduce and plateau to steady-state, suggesting two different friction dissipation mechanisms, as labelled in the figure. By examining the evolution of the structure during sliding, we discovered that the initial spike in shear force is due to the bridge that links the two membranes at the midplane. Figure S3 shows the time series of the $v$ = 0.25 m/s simulation with the structure, which demonstrates an intact bridge structure for the first 60 ns before its dissociation. During this period, while the lower bilayer is forced to follow the gold slab and travels more than one box length across the periodic boundary, the top bilayer moves only incrementally as indicated by the position of the pores and the marker lipid oxygen. Meanwhile, the lipid bridge has travelled significant distance following the lower membrane. This indicates that the bridge lipid bundle is being dragged through the lipids in the leaflet immediately above, and significant lipid tail-tail interaction arises, causing a surge in the shear stress measured. After the bridge dissociates, the bridge lipids retract back to the lamellar state and a friction dissipation is maintained at a lower shear stress about half of the peak values (Figure 3, Figure 4 a). This lower shear stress is attributed to a different friction dissipation mechanism related to the undulation of the membranes caused by the electropores, as we shall see next.

Is the bridge-related dissipation a relevant shear dissipation mechanism at SFB velocities, i.e. $O(\mu m/s)$, at steady-state? Like any bilayer, the bridge exposes hydrophilic groups and is stabilized by the hydrophobic interactions between the amphiphilic lipids which self-organized under thermal motion. As the two bilayers move relative to each other at decreasing velocities, the effect of lipid thermal motion, with characteristic timescale $\tau_{\text{diffusion}}$, becomes increasingly dominant over the instability introduced by lateral stretching, with characteristic timescale $\tau_{\text{translation}}$, and the bridge is more likely to be retained. Comparing these time scales of the two processes allows us to estimate the velocity limit below which the bridge will persist: $\tau_{\text{translation}}/\tau_{\text{diffusion}} = (L/v)/(L^2/D_{\text{lipid}}) = D_{\text{lipid}}/Lv \gg 1$, where $L$ is the characteristic length scale, $v$ is the velocity, and $D_{\text{lipid}}$ is lipid diffusivity. We arrive at $v \ll 0.005$ m/s with $D_{\text{lipid}} = 0.005$ nm$^2$/ns estimated from the simulations and $L = O(1$ nm$)$, the size of the bridge. In particular, at the much lower velocities of $O(\mu m/s)$ relevant to SFB experiments, the bridge structure remains stable and unperturbed by the sliding and plays an important role in friction dissipation in a similar way seen in our simulations before the bridges dissociate (Figure 3).

**Enhanced interbilayer headgroup attraction driven by electropore undulation.**
The increased shear stress measured with structure B ($n_w$=12, $E$=0.1 V/nm) relative to structure A ($n_w$=12, $E$=0 V/nm) in Figure 4 a reveals that, unexpectedly, the presence of electropores contributes to the enhanced frictional dissipation no less significantly than that of the inter-bilayer hydrophobic shear arising from the bridge formation. We attribute this as follows: Electropores introduce undulation in the membranes by their notable toroidal shape [28-30] and moreover, due to their sequestration of water, cause dehydration at the intermembrane space, which reduces the mean spacing between the opposing headgroups; this promotes increased inter-headgroup attraction. We assess the headgroup interactions using the distances between the phosphate (P) groups and the choline (N) groups from the opposing bilayer as a proxy. For each P or N atoms of the top midplane monolayer, we collected the distance to its nearest neighboring P or N atom in the bottom midplane monolayer and plotted the distribution (Figure S5 a).

Structure B, $n_w$=12, $E$≈0.1 V/nm, has proved to be the most headgroup interlocked system: the P-N distances has a significant peak at $r_{min}$=0.49 nm, which is roughly one water molecule size less than the next peak at 0.77 nm. The first peaks of the P-P and N-N pairs are found at 0.88 nm and 0.73 nm respectively. If we exploit the Lennard-Jones parameters assumed by the simulations: $\sigma_{PN}$≈0.36 nm $\sigma_{PP}$≈0.38 nm, and $\sigma_{NN}$≈0.33 nm, then the critical distances at the minimum potential energy can be estimated for the P-N, P-P, and N-N pairs as $2^{1/6}\,\sigma_{ij}$, giving 0.40 nm, 0.43 nm, and 0.37 nm. We then see that while the similarly charged P-P and N-N pairs are always separated by layers of water molecules, a fraction of the P-N pairs are in direct contact with each other with almost no water in between. The resulting + vs. – charge-charge attraction is then substantial.

The origin of the reduced inter-bilayer headgroup spacing is two-fold. Most apparently the inter-bilayer hydration level reduces during the formation of electropores due to sequestration of water within the pores. The reduced post-electroporation interbilayer hydration level is measured to be ca. 11 water/lipid. We test the dehydration effect by stripping water molecules from the $n_w$=12, $E$=0 system until $n_w$=11 and equilibrate again. The nearest neighbor test is shown in the inset of the bottom-left plot of Figure S5 a, and it is surprising to see that direct P-N coupling is not significantly promoted as seen in the yellow curve. This is because we overlooked the effect of enhanced undulation of the lipid membranes – electroporation introduces surface undulations manifesting as dimples at the pore region and humps in the opposing membrane. As the number of water molecules is fixed, undulation further stretches the midplane water phase thinner in terms of water molecules per area (Figure S5 c). To demonstrate this effect, we take the $n_w$=11, $E$=0 system and replace the solid surfaces with 6% increase in area, estimated from the lengths of the curved headgroup-headgroup interface as in Figure S5 b. The P-N distance analysis is shown in the same inset, and we see that the first peak is almost identical to that of the porated $n_w$=12 structure. In short, enhanced interbilayer P-N attraction is induced by electroporation-induced dehydration; this is consistent with the observation from a previous MD study, though without an external electric field, where such enhanced attraction was observed at $n_w$=4 [12].

**Shifted slip plane.**

In sum, the electric field promotes strong, hence more dissipative lipid-lipid interaction in series with hydration lubrication at the inter-bilayer midplane [1, 6, 31]. To understand the composition of the friction force, we propose a model based on three main mechanisms: direct inter-bilayer lipid interaction and hydration lubrication which act in series at the midplane, and the interleaflet interaction which acts in parallel to the total midplane shear stress. By taking the geometric center of the lipid system as the reference position, the velocities of the two midplane monolayers are approximately opposite and equal, which we denote as $v_s$ (Figure 4 c). The outer monolayers retain equal velocities $v_0$ of the solids to which they are effectively fixed. At steady-state, the monolayer velocities stabilize, and from force balance we can write:

$$\sigma = \frac{F}{A} = b(v_0 - v_s) = \eta \dot{\gamma} + \sigma_{\text{lipid}}$$

[1]

where $\eta$ is the dynamic viscosity of water, $\dot{\gamma}$ the shear rate across the midplane, and $\sigma_{\text{lipid}}$ the shear stress due to inter-bilayer lipid-lipid interactions. The interleaflet shear stress $b(v_0 - v_s)$ is a widely accepted model, where it is linearly dependent on the relative velocity with a scaling coefficient $b$ [14, 18]. The most suitable value of $b$ to our SFB experiments is O($10^{11}$) Pa s/m, estimated from the interleaflet friction measurements by Cao et al. (2021) [32]. Cao's experiment is performed under largely similar conditions, where lipid liposomes were compressed down to a single bilayer thickness with a pressure of ~10 atm, allowing the slip-plane to only occur between the leaflets (see Methods).

To evaluate the hydrodynamic term $\eta \dot{\gamma}$, the effective interbilayer dynamic viscosity of water $\eta$ is estimated from the Stokes-Einstein relation, which gives $\eta = \eta_0 \frac{D_0}{D}$, where $\eta_0$ and $D_0$ are respectively the dynamic viscosity and diffusivity of bulk water. The diffusivity of interbilayer water $D$ is evaluated from the slope of the linear fit to the mean square displacement in $y$-direction as a function of time (Figure S1). Despite the confinement effect, viscosity of water at the midplane remains fluid with at most 4 times of increase from the bulk viscosity of TIP3 water. This is reasonable as previous SFB studies demonstrated experimentally that the viscosity of water confined between charged mica surfaces is similar to bulk water down to subnanometer thickness [31, 33]. Shear rate $\dot{\gamma}$ is evaluated by taking the slope of the velocity profile $v(z)$ at the midplane (Figure S 1). Finally, we can evaluate $\sigma_{lipid}$ as a function of $v_s$ from Equation [ 1 ]; this is plotted in Figure 4 c. From the plot, we see that at small velocities comparable with those in the SFB ($v \sim \mu m/s$), $\sigma_{lipid}$ extrapolates to $O(10^5)$ Pa for the $E\approx 0.1V/nm$ systems, while roughly $\sigma_{lipid} \approx 0$ for $E=0$. At the experimental velocities $v=O(\mu m/s)$, the hydrodynamic stress is of $O(1)$ Pa ($\eta\dot{\gamma} = \frac{\eta v}{H} = \frac{10^{-3} \text{ Pa} \cdot 1 \text{ }\mu m/s}{1 \text{ nm}}= 1$ Pa, where $H$ is the thickness of the interbilayer water, taken to be ~1 nm, see Figure S 1). To simplify, we then reduce the right hand side of Equation [ 1 ] as $\eta\dot{\gamma}$ and $\sigma_{lipid}$ respectively for $E=0$ and $E\approx 0.1V/nm$. The velocity of the inner monolayers $v_s$ is usually unknown in experiments, and we can reduce it by rearranging the equation to get $v_s = v_0 - \frac{\eta\dot{\gamma}}{b}$ and $v_s = v_0 - \frac{\sigma_{lipid}}{b}$ respectively for $E=0$ and $E\approx 0.1V/nm$. We can then write

$$E = 0: \quad \frac{v_s}{v_0} = 1 - \frac{\eta\dot{\gamma}}{bv_0}$$

$$E \approx 0.1 \text{ V/nm}: \quad \frac{v_s}{v_0} = 1 - \frac{\sigma_{lipid}}{bv_0} \quad (\sigma_{lipid} < bv_0)$$

[ 2 ]

At experimental conditions, $b\sim O(10^{11})$ Pa s/m, $v_0 \sim O(10^{-6})$ m/s, $\eta\sim 10^{-3}$ Pa m/s, and $\dot{\gamma}\sim \frac{v_0}{H}\sim 10^3 s^{-1}$. Substituting $b = O(10^{11})$ Pa s/m as previously discussed, we acquired for $E=0$, $\frac{v_s}{v_0} \approx 1$, which suggests minimal interleaflet shearing, and dissipation mainly occurs at the interbilayer water phase. For $E\approx 0.1V/nm$, since $\sigma_{lipid}$ and $bv_0$ are of similar orders of magnitude at $O(10^5)$ Pa, either $\frac{v_s}{v_0}\sim O(0.1)$ if $\sigma_{lipid} < bv_0$, where dissipation occurs at both midplane and the interleaflet planes, or $\sigma_{lipid} > bv_0$ and no midplane sliding occurs at all. We then conclude that phenomenologically in the experiment, when we switch on the electric field, the slip plane shifts from the midplane, where hydration lubrication was dominant, to the interleaflet planes, partly if not completely.

## 3    Conclusion

In this work, we elucidated the molecular mechanism of a reversible and in-situ method of massive electro-tuning of friction between surfaces coated with hydrated phosphatidylcholine (PC) lipid bilayers, as discovered in a recent SFB study [13]. Our previous all-atom molecular dynamics (MD) simulations showed that an electric field of 0.1 V/nm, similar to that applied in the SFB experiments induces topological convolution at the membrane-membrane interface through electroporation, which significantly alters the interaction between the two lipid bilayers [21]. These structural changes prevent the membrane's shearing friction from being dissipated through the hydration lubrication mechanism only. The electric field induces pores in the lipid bilayer, causing some dehydration at the inter-bilayer region and enhanced headgroup attraction between the oppositely charged phosphate groups and choline groups which acts to increase the dissipation as they slide past each other. This effect is prevalent for varying hydration levels and velocities and is only dependent on the presence of pores. Electroporation also promotes stochastic lipid motion, which increases the rate of a formation of intermembrane hydrophobic contacts. This leads to the second effect – formation of lipid bridges, which occurs sporadically and is associated with locally excess hydration from the water defects naturally found in supported lipid bilayers [34, 35]. Such bridges are stable at the experimental sliding velocities, forcing slip between the surfaces to occur at the acyl tail interface associated with higher friction. For both mechanisms, the effect of the electric field is to enhance frictional dissipation by lipid-lipid interactions at the midplane. The shear stress incurred, as extrapolated from our simulation results to experimental sliding velocities, $\sigma_{\text{lipid}} \sim O(10^5)$ Pa, is similar to the interleaflet friction stress measured experimentally under similar conditions [32]. This implies that the slip plane concurrently occurs between the interleaflet planes in addition to the midplane.

Using molecular dynamic simulations to study friction has always been challenging due to the limitation in time and length scales. Here we present a case where extensive precaution was taken for modeling and protocol design, which enabled us to relate molecular details to macroscopic measurements in a quantitative way. From the simulations, the friction coefficient within a PC lipid membrane stack exhibits a noticeable increase from almost zero to 0.1 in response to an electric field stimulus of 0.1 V/nm, which is in excellent agreement with the SFB measurements. Our results shed further lights on designing hydration-lubricants and developing electro-tuning applications through fast sampling afforded by molecular dynamics predictions.

## Supplementary Materials

### Videos

Video 1: Time series of $n_w$=20, E=0.1 V/nm, v = 0.25 m/s. The orange spheres are lipid headgroup oxygen O21 atoms used to indicate the positions of the leaflets and pores. Selective oxygen atoms are labelled blue for visual cues of each leaflet's motion.

Videos 2-3: The dynamic interaction of the headgroups during the shearing motion at $v$=0.1 m/s for both the $n_w$=12 and the $n_w$=20 systems respectively.

### Methods

**Membrane stacks equilibrated under electric field**. Please refer to our previous work for full description of method [21].

**Shearing simulations of membranes with planar solid confinement.** The cross-section area is fixed under the NVT ensemble. As described in our previous work, the electric field is maintained using the charge-imbalance method by transferring naturally dissociable potassium ions from the mica surface to the gold surface [36-47]. The mica surface is fixed in all three directions using the freeze group option. The potassium ions on the gold surface are restrained in $z$. The y coordinates of the gold surface are fixed to prevent lateral sliding. In addition, for the case of $n_w$=20, nitrogen atoms of the lipid monolayer in immediate vicinity of 1 nm range to the mica surface are fixed. This is to correct for the unrealistic effect that the excess water molecules presented reduces lipid-mica interaction. Liposomes are known to interact strongly with cleaved mica surfaces. They tenaciously cling to the mica surface even when flushed by a flow of water at roughly 1 m/s [6]. A normal constant compression force of 53 kJ mol$^{-1}$nm$^{-1}$ ($\approx$ 88 pN) is applied between the solid slabs. A harmonic potential $U = \frac{k}{2}(vt - x(t))^2$ with a force constant of $k$ = 1000 kJ/(mol·nm$^2$) was applied to the center of mass of the gold slab, pulling the gold slab at constant velocities of 0.1 m/s, 0.25 m/s, 0.5 m/s, 0.75 m/s, and 1 m/s in the $x$-direction. The resulting shear force is then evaluated as $F_f = -\frac{dU}{dx} = k(vt - x(t))$. Shear stress $\sigma = F_f/A$, where $A$ is the cross-section area of the systems.

**Estimating the number of splayed lipids.** Coordinates of the end carbon atoms and the oxygen atom connecting the two acyl tails are exported using GROMACS functions from the simulations and analyzed in MATLAB. A splayed lipid is identified if the vectors pointing from the oxygen atom to the two carbon atoms have different signs and each with magnitude longer than 1 nm. The $n_{splayed}(t)$ curve function is then smoothed over 5 ns intervals.

**Interbilayer headgroup-headgroup distance analysis.** The distribution of distances to the nearest neighbor for headgroups of the two inner monolayers were evaluated as the following: $d_{\min} = \min{(\vec{i}_n - \vec{j}_m)}$ ($m=1...M$) are collected for each $n$ in $1..N$, where $\vec{i}$ and $\vec{j}$ are permutations of $\vec{P}$ and $\vec{N}$, $n$ and $m$ are the indices of headgroups of the two midplane monolayers. A distribution function is then calculated from the set of $d_{\min}$.

**Analysis of dynamics.** From trajectories of the shearing simulations, velocities of water and lipids are evaluated by slices in $z$ and averaged through time during the steady-state stage. Lateral diffusivities are evaluated from the ratio of the mean square displacement in $y$ and the time interval. Mass density of water and number density of lipid carbon (C35) atoms are plotted as a function of $z$. The diffusivities and velocities for each lipid monolayer are extracted at the $z$ coordinates of the four number density peaks. Diffusivity of the interbilayer water phase is extracted at the $z$ coordinate of the mid-plane mass density peak. Shear rate is determined by taking the gradient of the velocity profile near the same $z$ position (Figure S 1*)*.

**Evaluation of the interleaflet friction coefficient $b$.** The $b$ coefficient itself is largely dependent on the sliding velocity and lipid species [17, 18]. Experimentally, the range of $b$ of various types of phosphocholine lipids lies within $10^7$-$10^9$ Pa s/m, measured by methods such as lipid diffusion or lipids' response to a microchannel shear flow, dissimilar to the SFB experimental conditions [18, 48, 49]. By existing MD simulation studies, the magnitude is $10^6$ -$10^7$ Pa s/m at velocities of O(m/s), and pronounced shear-thinning behavior is observed [14, 17]. We can estimate the value of $b$ from our simulations as $b = \sigma/(v_0 - v_s)$, which gives a range of 1- $8 \times 10^7$ Pa s/m. Our value is slightly higher than existing MD studies of lipid-only systems because lipid thermal motion is restrained by the solids with reduced lipid diffusivities [18]. To evaluate the interleaflet friction in our SFB experiments, perhaps the most representable value of $b$ is O($10^{11}$) Pa s/m, estimated from the interleaflet friction measurements by Cao et al. (2021) [32]. In Cao's experiment, phospholipid liposomes were compressed down to a single bilayer thickness with a pressure of ~1 MPa, and the friction coefficient at this pressure is measured to be O(0.1) at a shearing velocity of 0.925 μm/s. It should not be surprising that this $b$ value is a few orders of magnitude higher than our simulated values considering the effect of shear-thinning — velocities in SFB experiments are 6 orders of magnitude lower than simulated velocities at O(0.1) m/s.

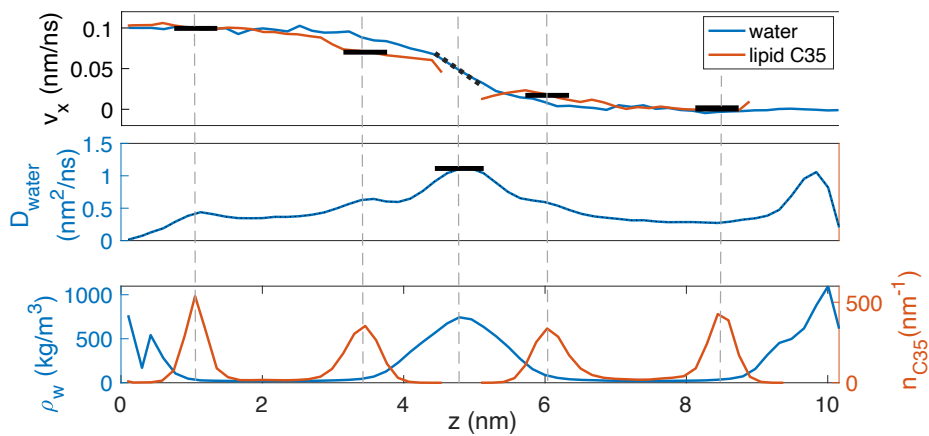

Figure S 1. Analysis of dynamics with shearing simulation of $n_w$=12, $E \approx 0.1$ V/nm. Top: velocity in the $x$-direction of water and lipid evaluated by averaging $v_x$ over time and over atoms fall into each slice in $z$. Middle: diffusion coefficient of water in the $y$-direction. Displacements over a finite time step $dt$ of atoms fall into each slice in z are collected. Lateral diffusivities are then evaluated as $\langle (y(t+dt) - y(t))^2 \rangle / dt$. Bottom: water mass density and number density of lipid C35 atoms as a function of $z$. Representative values of $v_x$ and $D$ (black bar) are extracted at the z-coordinates of the mass peaks (grey dashed lines).

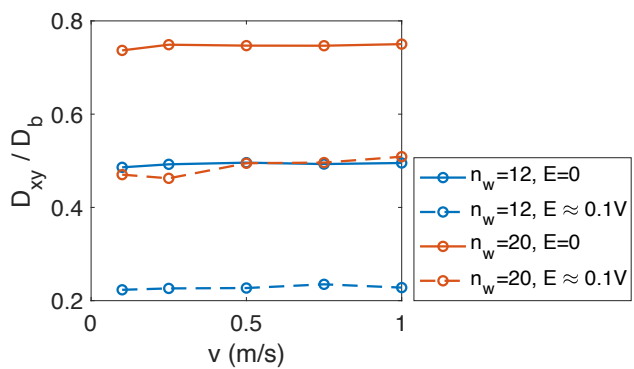

Figure S 1. Lateral diffusivity of water at the mid-plane for all four systems investigated. The bulk water self-diffusivity of TIP3P is $D_b$=5 nm²/nm [50, 51].

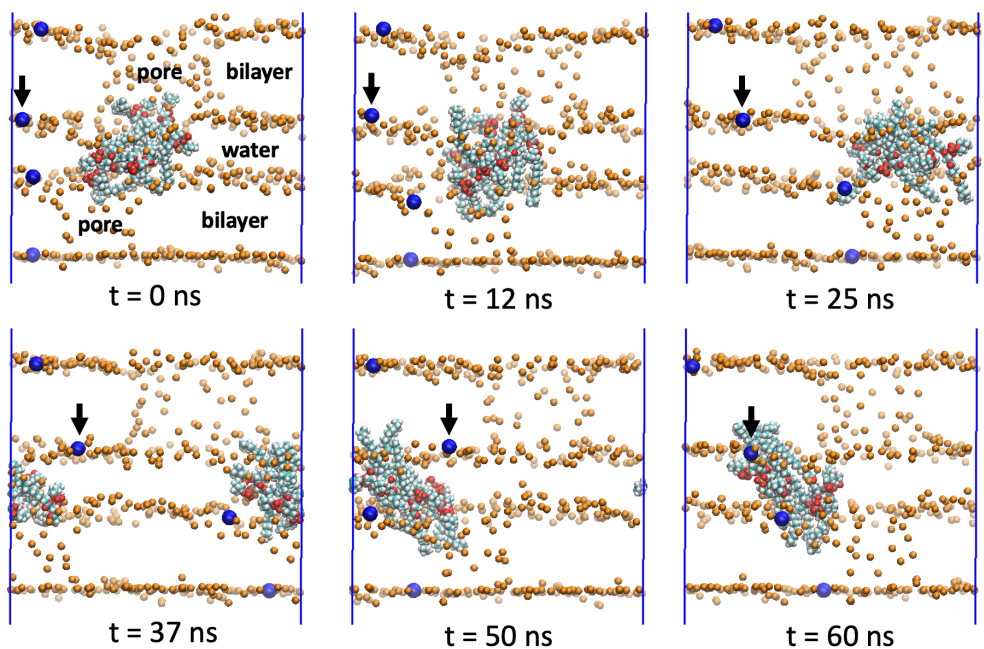

Figure S3. Time series of $n_w$=20, $E$=0.1 V/nm, $v$ = 0.25 m/s. The bridge structure remains intact for 60 ns and dominates the friction dissipation at the midplane. The orange spheres are lipid headgroup oxygen O21 atoms used to indicate the positions of the leaflets and pores. Selective oxygen atoms are labelled blue for visual cues of each leaflet's motion. The arrows demonstrate that the top bilayer only moves incrementally while the bridge and the lower bilayer have travelled a box length. Positions of the two electropores are another visual cue for the same conclusion. Also see Video 1 in SI.

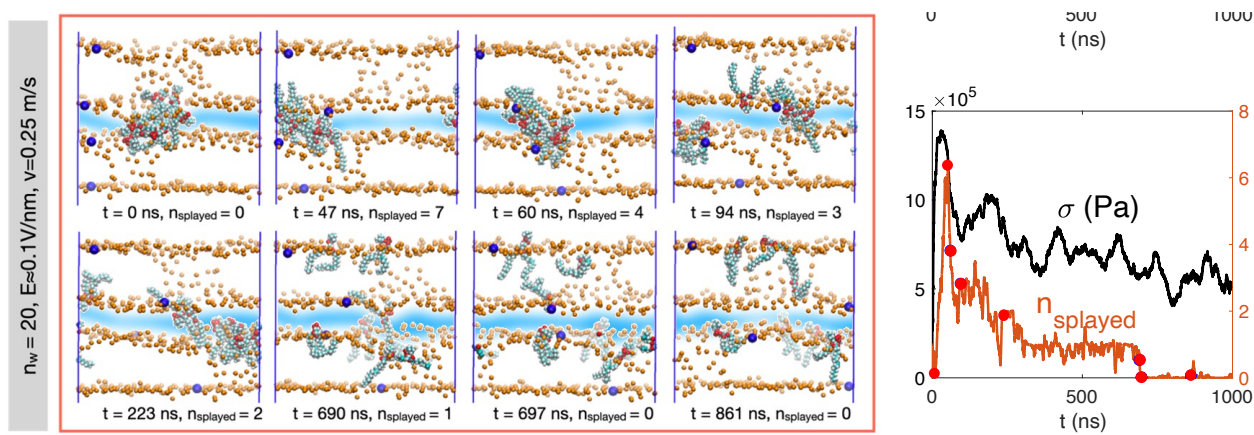

Figure S4. Time series of $n_w$=20, $E$=0.1 V/nm, $v$ = 0.25 m/s. Top: Correlation of number of inter-membrane lipid molecules and adopted from Ref. [52].

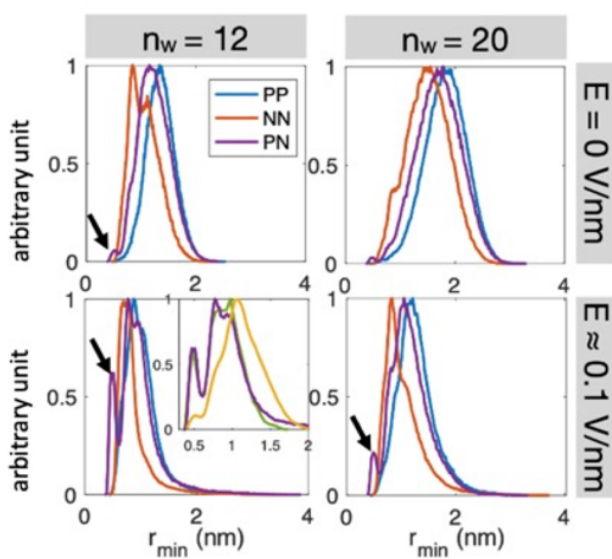

Figure S5. Distributions of the distances to the nearest neighboring headgroup atoms in the opposing midplane monolayers. Inset in bottom-left: purple, yellow, and green respectively corresponds to the distribution of P-N pair from the $n_w$=12 and $E$≈0.1 V/nm system, from the $n_w$=11 and $E$=0 V/nm system with the same cross-section area, and $n_w$=11 and $E$=0 V/nm system with cross-section area of 6% increase. The arrows indicate $r_{min}$=0.49 nm where P and N are in direct contact with no water in between.

| Reference | Fold-change (method/$\mu_{low}$/$\mu_{high}$) | Reversibility |
|---|---|---|
| Phys Rev Lett 2012, 109 (15), 155502. | 3.5 (colloidal tip AFM/0.16/0.56) | Reversible |
| *Phys Rev Lett 2012, 109 (15), 154302. | ~8.5 (SFB/0.05/0.006) | Reversible |
| Adv Mater 2013, 25 (15), 2181-2185. | Only measure adhesion, 5 folds | Reversible |
| Langmuir 2009, 25 (20), 12114-12119. | 3 (colloidal tip AFM) | Reversible |
| Tribology Letters 2013, 53 (1), 17-26. | 7 (tribometer/0.02/0.15) | Reversible |
| Tribology Letters 2010, 38 (2), 169-178. | 6.4 (tribometer/0.07/0.45) | Reversible |
| Langmuir 2011, 27 (6), 2561-6. | 10 (AFM/0.04/0.4) | Reversible |
| Science 2006, 313 (5784), 186. | 3.1 (AFM/0.08/0.25) | Reversible |
| Chem Commun (Camb) 2014, 50 (33), 4368-4370. | 19 (AFM/0.019/0.001) | Not mentioned |
| Journal of The Electrochemical Society 1992, 139 (12), 3489-3492. | **2.1 (tribometer/0.25/0.52) | Reversible |
| Langmuir 2018, 34 (3), 801-806. | 26 (colloidal tip AFM/0.12/3.1) | Reversible |
| Phys Chem Chem Phys 2013, 15 (35), 14616-23. | 4 (colloidal tip AFM/0.125/0.5) | Irreversible |
| The Journal of Physical Chemistry C 2018, 122 (39), 22549-22555. | 3.6 (colloidal tip AFM/0.28/1) | Reversible |
| Soft Matter 2019, 15 (21), 4255-4265. | Not mentioned, only Fn is measured | Reversible |
| The Journal of Physical Chemistry C 2020, 124 (43), 23745-23751. | 2 (colloidal AFM/0.0031/0.0065) | Reversible |
| Tribology Letters 2021, 69 (4). | 4 (tribometer/0.1/0.4) | Reversible |
| Tribology Letters 2010, 41 (3), 485-494. | 9 (tribometer/0.05/0.45) | Irreversible |
| Journal of Materials Chemistry C 2017, 5 (24), 5877-5881. | 2.4 (tribometer/0.14/0.24) | Reversible |

Table 1. Existing literature on electro-tuning mechanism. Note that 200-fold electro-modulation achieved by the current study is one to two orders of magnitude larger than any external friction modulation reported to date.